\documentclass[manuscript]{aastex}

\usepackage{xcolor}

\fullcollaborationName{The Friends of AASTeX Collaboration}

\begin{document}

\title{The radial variation of the solar wind turbulence spectra near the kinetic break scale from Parker Solar Probe measurements}

\author{S. Lotz\altaffilmark{1,2,3}, A.~E. Nel\altaffilmark{1,2}, R.~T. Wicks\altaffilmark{1,4}, O.~W. Roberts\altaffilmark{5}, N.~E. Engelbrecht\altaffilmark{6,7}, R.~D. Strauss\altaffilmark{6,7}, G.~J.~J. Botha\altaffilmark{4}, E.~P. Kontar\altaffilmark{8}, A. Pit\u{n}a\altaffilmark{9},  S.~D. Bale\altaffilmark{10}}



\altaffiltext{1}{Equal contribution authors}
\altaffiltext{2}{South African National Space Agency, Hermanus, 7200, South Africa}
\altaffiltext{3}{MUST, Faculty of Engineering, North-West University, South Africa}
\altaffiltext{4}{Department of Mathematics, Physics and Electrical Engineering,  Northumbria University, Newcastle upon Tyne, NE1 8ST, UK}
\altaffiltext{5}{Space Research Institute, Austrian Academy of Sciences, Schmiedlstrasse 6, Graz, 8042, Austria}
\altaffiltext{6}{Center for Space Research, North-West University, Potchefstroom, 2522, South Africa}
\altaffiltext{7}{National Institute for Theoretical and Computational Sciences (NITheCS), South Africa}
\altaffiltext{8}{School of Physics and Astronomy, University of Glasgow, Glasgow, G12 8QQ, UK}
\altaffiltext{9}{Department of Surface and Plasma Science, Faculty of Mathematics and Physics, Charles University, V Hole\u{s}ovi\u{c}k\'ach 2, 180 00 Prague, Czech Republic}
\altaffiltext{10}{University of California, Berkeley, USA}

\begin{abstract}
In this study we examine the radial dependence of the inertial and dissipation range indices, as well as the spectral break separating the inertial and dissipation range in power density spectra of interplanetary magnetic field fluctuations using {\it Parker Solar Probe} data from the fifth solar encounter between $\sim$0.1 and $\sim$0.7 au. 
The derived break wavenumber compares reasonably well with previous estimates at larger radial distances and is consistent with gyro-resonant damping of Alfv\'enic fluctuations by thermal protons. 
We find that the inertial scale power law index varies between approximately -1.65 and -1.45.
This is consistent with either the Kolmogorov (-5/3) or Iroshnikov-Kraichnan (-3/2) values, has a very weak radial dependence with a possible hint that the spectrum becomes steeper closer to the Sun.
The dissipation range power law index, however, has a clear dependence on radial distance (and turbulence age), decreasing from -3 near 0.7 au (4 days) to -4 [$\pm$0.3] at 0.1 au (0.75 days) closer to the Sun.

\end{abstract}



\section{Introduction}

High-cadence magnetic field observations made by the Parker Solar Probe (PSP) during its various perihelia provide us with an unprecedented opportunity to study the radial evolution of various quantities associated with the dissipation of solar wind turbulence \cite[see, e.g.,][]{woodham2019,PerroneEA21,ChhiberEA21,AlexandrovaEA21}, motivated by the fact that a greater understanding of the processes of kinetic dissipation in magnetized plasma is essential for explaining the physical origin and evolution of the solar wind \cite[e.g.][]{BrunoCarbone16,Matt20,Duan2020,TelloniEA21,Zhao22}. This, in turn, also informs modelling efforts of turbulence transport \cite[e.g.][]{Eng2018,ChhiberEA19,AdhikariEA20,AdhikariEA21} and energetic particle transport \cite[e.g.][]{StraussEA17,Eng19,LD19,ChhibberEA21b} and relates to density turbulence that is important for solar radio burst interpretation \citep[e.g.][]{krupar20, kontar19}. Magnetic field fluctuations in the solar wind are commonly observed to follow a power-law spectrum. The inertial range is created by an energy-conserving spectral cascade \cite[e.g.][]{smith06} where interactions between fluctuations can still be described by fluid dynamics \cite[however, for an exception, see][]{bian10}. 
The spectral index in the inertial range has been observed to be close to both the \cite{kol41} value of $-5/3$, as well as the Iroshnikov-Kraichnan value of $-3/2$ \cite[see, e.g.,][]{smith06,Podesta11,ZhaoEA20,chenetal2020}. 
This range is followed by a break and then a steepening in the magnetic field power spectrum, where the MHD description breaks down and kinetic effects of individual particles, and thermal heating, start playing a role \citep{AlexandrovaEA08}. 
The dissipation range spectral index is dependent on the type of turbulent fluctuation present, being either Alfv\'en waves or coherent structures, and can vary significantly, ranging between $\sim -1$ and $-4$ \citep{LeamonEA98b,smith06,MarkovskiiEA06,Alexandrova2008,BrunoEA14,Bruno2014,Lionetal2016,VechEA18,franci20}.\\

The physical mechanisms responsible for the break between the inertial and dissipation range at these ion-kinetic scales are still 
not fully understood~\citep{Matt20}. 
The processes at play in energy transport in the transition range and how it affects the properties of plasma is still an open question~ \cite[e.g.][]{GoldsteinEA15,BrunoCarbone16,TerresLi22}.
The break frequency $f_b$, where the transition from inertial to dissipation range is observed, has been shown to display a radial dependence in previous studies. 
Analyzing observations taken in the 0.42 to 5.3 au range during radial alignments respectively between MESSENGER and WIND, and WIND and ULYSSES, \cite{Bruno2014} found that the break frequency increased as the heliocentric distance decreased such that $f_b \propto r^{-1.09 \pm 0.11}$. 
In another study \cite{Duan2020} used data from the cruise phase of the second orbit of PSP, ranging from 0.17 to 0.63 au, to measure the spectral break that those authors interpret as the transition to kinetic turbulence. 
They found that the break frequency increased with a decrease in heliocentric distance, following a power law of $f_b \propto r^{-1.11 \pm 0.01}$. \\

A wide array of studies examine time periods for the break scale and find evidence for a link to the kinetic plasma physics of protons~ \citep[e.g.][]{GoldsteinEA94,HamiltonEA08,MarkovskiiEA08,SmithEA12,Chenetal2014,WoodhamEA18}. 
At 1 au the ion scale spectral break tends to occur near $f\sim 0.1-1$~Hz in the spacecraft frame \citep[e.g.][]{LeamonEA98b,Smith2001,Bale2005,MarkovskiiEA08,Bourouaine2012}. 
The break is located near the spacecraft frame frequencies that correspond to either the proton gyroradius $l_g$ where damping of kinetic Alfv\'en waves becomes significant or the proton inertial length $l_i$ where protons decouple from the turbulent magnetic field. 
It may also be associated with magnetic reconnection via the so-called disruption scale \cite[for more detail see][and references therein]{VechEA18,TerresLi22}. 
The break frequency has been observationally associated with these various characteristic plasma lengthscales, for example \citet{LeamonEA98b} and \citet{LeamonEA00} report from analyses of {\it WIND} data that this break occurs at spatial scales in the plasma frame near the proton gyroradius. Alternatively the break may be related to the combined scale $l_i + l_g = 2\pi k_c^{-1}$, which is associated with cyclotron resonance of Alfv\'en waves propagating along the mean field direction \citep[e.g][]{LeamonEA98b,Bruno2014,WoodhamEA18,Eng2018}. At 1 au it is often the case that  plasma $\beta \sim 1$ implying $l_g \sim l_i$, which makes it difficult to determine which of spatial scales is related to the break \cite[see, e.g.,][]{TerresLi22}. To overcome this limitation, \cite{Chenetal2014} investigated intervals with extreme values of plasma $\beta$ where the two scales are well separated. It was found that the break tends to be associated with the larger of the two scales, which is consistent with the break being near the combined scale. \cite{Bruno2014} performed a study of how the spectral break changes with heliocentric distance and found the best agreement to be with the combined scale, with a radial dependence of the wavenumber at which this break occurs of $k_b \propto r^{-1.08 \pm 0.08}$. A study of Wind data also confirmed that the best agreement is with the combined scale \cite{WoodhamEA18}.\\

In this work we determine the spectrum near the break wavenumber and spectral indices using high-cadence PSP magnetic field measurements taken during its fifth orbit, in an extension of the study of \citet{Duan2020}. 
The fifth orbit includes observations from 07.05.2020 to 19.06.2020, with PSP radial distance from the Sun varying between 0.1 and 0.7 au.
These values are then compared to the theoretical estimates for this quantity by employing in situ observations for the various plasma quantities they are a function of in an attempt to determine which is the dominant turbulence dissipation process in the inner heliosphere. 
The break wavenumber is then quantitatively compared to previous studies and the radial dependence across the widest range of heliocentric distances yet measured is calculated. 
Furthermore, the radial evolution of the inertial and dissipation range spectral indices are also investigated. 
The next section details the analysis method employed in this study.
Sections~\ref{sec-kd} and \ref{sec-si} present the results of this study with regards to the dissipation range onset frequency/wavenumber, and spectral indices, respectively. 
Finally, the results are discussed in Section \ref{sec-discussion}.

\section{Data Analysis} 

\begin{figure}
    \centering
    \includegraphics[width=0.99\textwidth]{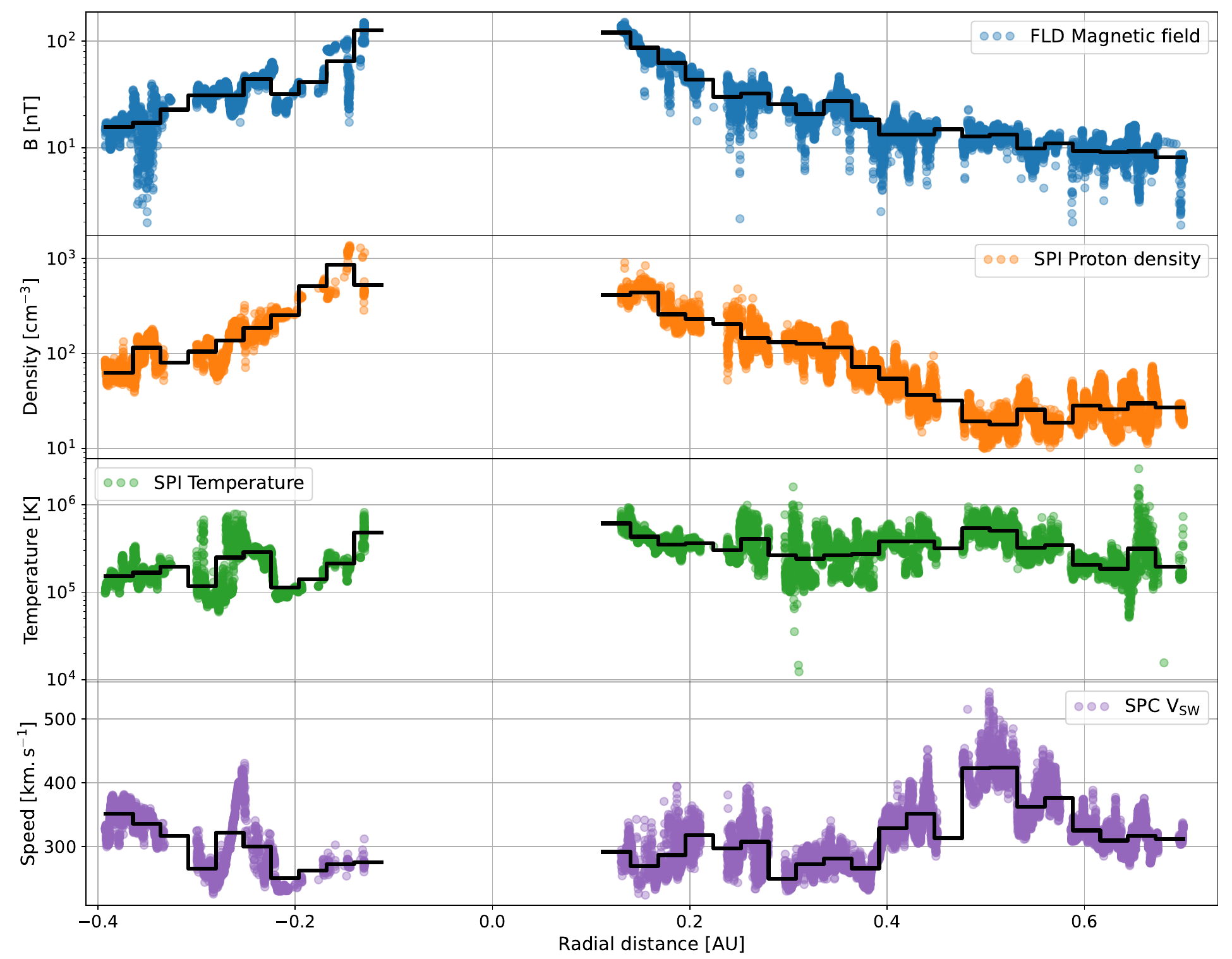}
    \caption{An overview of the solar wind plasma measurements during PSP orbit 5. 
            {The black stepped curves indicate the local mean for 50 equispaced radial distance bins.}
            }
    \label{fig:PSP_plasma_overview}
\end{figure}

PSP is a 3-axis-stabilized Sun-pointing spacecraft in a elliptical heliocentric orbit, with aphelia between Earth and Venus~\citep{whit20}. 
The Solar Wind Electrons, Alphas, and Protons (SWEAP) instrument suite onboard the PSP primarily measures solar wind thermal plasma. 
The suite consists of three Electrostatic Analyzer instruments, called the Solar Probe ANalyzers (SPANs): A sun-pointing Faraday Cup (SPC) that primarily measures protons, alpha particles, and periodically electrons and the Solar Probe Analysers (SPAN) that are situated at either side of the spacecraft bus and measures protons, alpha particles, heavy ions (SPAN-Ion), and electrons (SPAN-Electron). 
We use density and thermal speed data from SPC and temperature from the SPAN-Ion instrument. 
These instruments were designed to overlap their fields of view and capabilities, and make complementary measurements \citep{sweap}. 
{We restricted the analysis to periods when data quality (quantified by the quality flag parameters) was at it's highest level for the SPAN, SPC and MAG instruments.}
The three-component heliocentric ($R$, $T$ and $N$ where $R$ is in the radial direction, $T$ is perpendicular to $R$ and lying in the equatorial place, and $N$ is normal to this plane and completes the right-handed coordinate system) solar wind magnetic field is measured by the MAG fluxgate magnetometer from the FIELDS instrument \citep{Bale2016}. 
During this period the spacecraft moved, in terms of radial distance, between $\sim$ 0.1 and 0.7 au. 
Fig. \ref{fig:PSP_plasma_overview} shows, from top to bottom, the magnetic field magnitude, the solar wind proton density, the effective temperature, and the solar wind speed against radial distance for the inward and the outward part of the orbit; note the logarithmic y-axis necessary to capture the wide change in parameters over the orbit in the top three panels.

\begin{figure}[]
    \centering
    \includegraphics[width=0.99\textwidth]{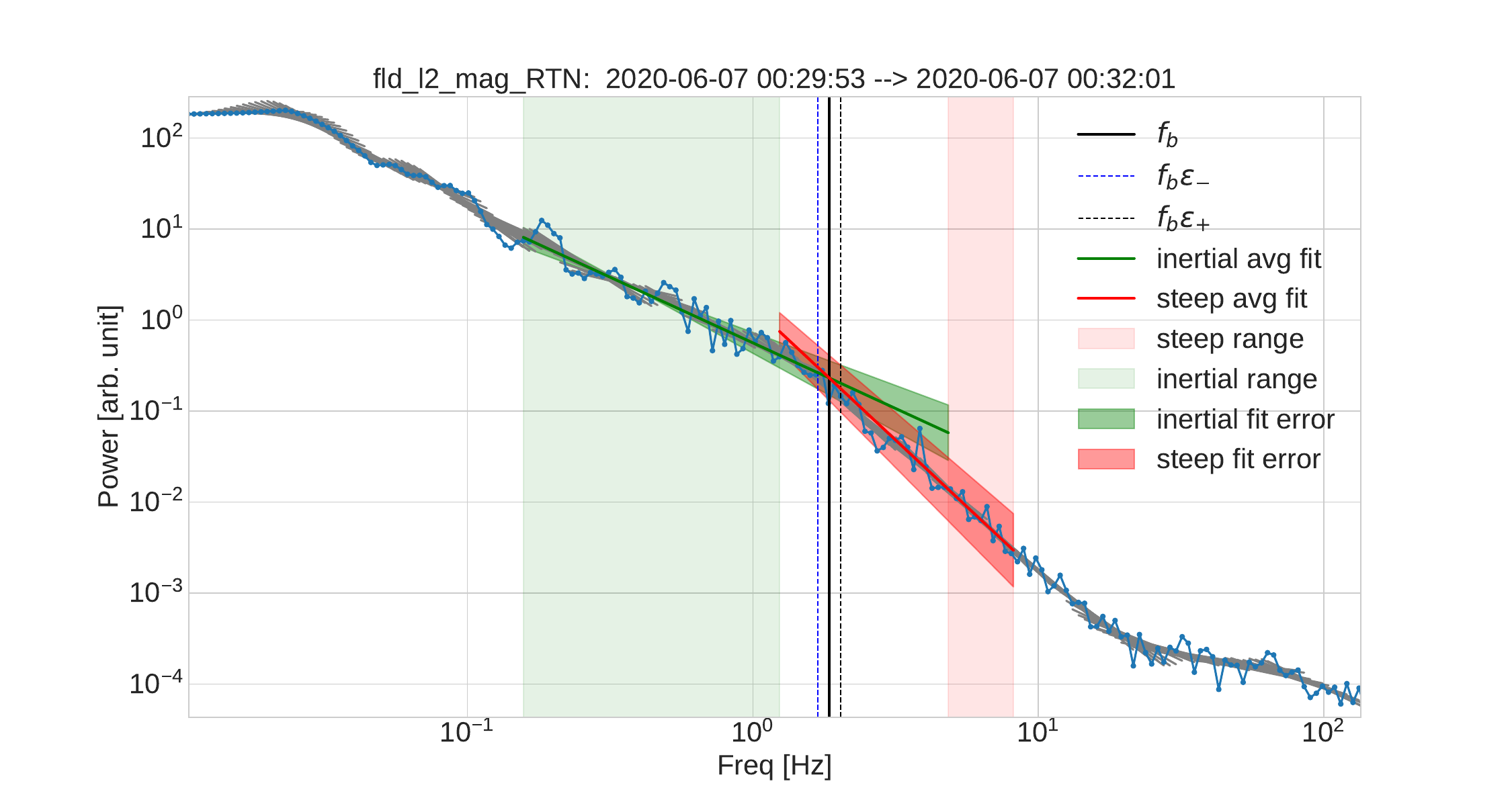}
    \caption{Magnetic field spectrum from a solar wind interval measured by PSP. Blue dotted line: initial spectrum. Grey lines: individual slopes measured over 20 points, and used to determine the 1) green slope which is the average fit for the inertial range, and 2) red slope which is the average fit for the dissipation range. The vertical black line represents the break frequency approximation. The green highlighted section defines the cutoff for the inertial range estimation and the red-highlighted section the cutoff for the dissipation range estimation.}
    \label{fig:breakfreq}
\end{figure}

The magnetic field is observed at data rate $d$ quantified by samples per instrument cycle.
For this study we only utilised intervals with data rate $d \ge 64$ samples / cycle.
The instrument cycle duration is $L \approx 0.874$s \citep{Bale2016}, i.e. $d=64$ would translate to 55 samples per second.
The time between samplings is not always consistent and to rectify this we interpolate the $B$ time series data to a fixed rate of $L/d$. 
We estimate the break frequency in the magnetic field spectra ($f_b$) using a procedure similar to the method employed by \cite{woodham2019}.
The time series spanning 44 days is divided into non-overlapping windows of 128s each, and $f_b$ is estimated for each of these windows.
The approximately 1025 hours of Orbit 5 equates to roughly 28,000 128-second intervals; however only a fraction of these intervals results in reliable $f_b$ estimates.
The $f_b$ estimation procedure is illustrated in Figure \ref{fig:breakfreq} for one such interval.

For every 128 second interval of $\mathbf B = [B_R, B_T, B_N]$ ($N \times 3$ matrix with $N$ the number of magnetic field observations) the power spectral density of $\mathbf B$ is computed by taking the trace $ P(f) = tr\lbrace \mathbf P(f) \rbrace$  of the matrix $\mathbf P(f) = \tilde{ \mathbf B}(f) \tilde{\mathbf B}(f)^*$, where $\tilde{\mathbf B}(f)$ denotes the Fourier components of $\mathbf B$ at frequency $f$ and $\tilde {\mathbf B}(f)^*$ its complex conjugate.
Since the shape of the spectral break is more pronounced in log space than linear space we linearly interpolate the power terms along log-spaced frequencies, denoted $P_l(f_l)$.
This is the blue dotted curve in Figure \ref{fig:breakfreq}.
Overlapping linear fits are applied to $P_l(f_l)$ and the average slope and intercept are determined within the inertial and dissipation ranges (green and red shaded regions in Fig. \ref{fig:breakfreq}).
These average fits are indicated by the red and green lines and their intersection determines the break frequency estimate $f_b$ (black vertical dashed line in Fig. \ref{fig:breakfreq}).
The mean standard deviation of the linear fits are used to estimate the uncertainty, denoted by the shaded regions around the average linear fits.
We define the uncertainty in the $f_b$ estimate as the frequencies where the upper and lower bounds of the linear fit uncertainty regions overlap (see Fig. \ref{fig:breakfreq}).

This yields a set of 26,581 estimates of $f_b$ along with an upper and lower uncertainty ($f_b \epsilon^-$, $f_b \epsilon^+$) for the orbit 5 data set.
The uncertainty estimates allow us to discard problematic $f_b$ estimates.
All estimates with (i) $f_b \epsilon^- > f_b \epsilon^+$, or (ii) $f_b$ outside the range of frequencies in the signal, or (iii) $f_b \epsilon^-$ or $f_b \epsilon^+$ beyond the dissipation and inertial ranges are discarded.
This results in a final set of 8551 reliable estimates of $f_b$. 


From the estimated break frequency, and the measured solar wind speed in each interval, a corresponding wavenumber is calculated as
\begin{equation}
 \label{Eq:kd_from_data}
     k_b =  \frac{2 \pi f_b}{V_{sw}},
 \end{equation}
with associated break scale $l_d = 2 \pi/k_b$. {As such, the Taylor hypothesis is explicitly used in this study. For turbulence analyses performed on observations taken specifically near the Alfv\'en critical point, this could be problematic \cite[see, e.g.,][]{BP18,BP19,BP20}. In terms of the dispersive regime specifically, \cite{HowesEA14} report that such a regime, were it primarily supported by Whistler waves, would violate Taylor's hypothesis, while \citet{KleinEA14} expect a flattening of the dissipation range due to this issue.} Note also that we do not take the angle between the flow and magnetic fields into account \citep{Chenetal2014,Bourouaine2012,DuanEA18}. 

In order to compare our results with the various proposed break wavenumbers, these quantities are calculated directly from PSP observations. 

{As a first approach, temperature isotropy ($T_{\text{eff}} = T_{||} = T_{\perp}$) is assumed,} allowing the effective plasma proton temperature to be calculated from the equipartition theorem as $m u_{\text{eff}}^2 = 3 kT_{\text{eff}}$, where $u_{\text{eff}}$ is the effective thermal speed.
From this, the proton gyro-scale can be estimated as
\begin{equation}
\label{Eq:gyroscale}
    l_g = 2 \pi \frac{u_{\text{eff}}}{\Omega_{ci}}  = \frac{2 \pi}{k_g} ,
\end{equation}
with  $\Omega_{ci}=qB/m_{i}$ the proton gyro-frequency. The proton inertial length now follows as
\begin{equation}
    l_i = 2 \pi\frac{c}{\omega_{ci}} =  2 \pi \frac{V_A}{\Omega_{ci}} = \frac{2 \pi}{k_i} ,
\end{equation}
with $V_{A}=B/\sqrt{\mu_{0} n_{i}m_{i}}$ the Alfv\'en speed. 
Thermal particles can also resonate with circularly polarized waves when, in the guiding centre frame, the following resonance condition is met
\begin{equation}
    \omega^* = n \Omega^*,
\end{equation}
where $\omega^*$ is the wave frequency, $\Omega^*$ the particle cyclotron frequency, and $n = \pm 1$ labels left-hand and right-handed waves, respectively. Transforming back to the bulk flow frame, the resonance condition becomes
\begin{equation}
\omega - \vec{k} \cdot \vec{v}  = n \Omega.
\end{equation}
Assuming left-handed, parallel propagating Alfv\'en waves with $\omega^{2}=k_{\parallel}^2 V_{A}^2$, resonating with thermal protons, Doppler shifted by their (parallel) thermal speeds $u_{\text{eff}}$, one obtains the so-called proton cyclotron resonance as \cite[see, e.g.,][and references therein]{Eng2018}
\begin{equation}
\label{Eq:cyclotron_resonance}
    k_c = \frac{\Omega_{ci}}{V_A + u_{\text{eff}}} = \frac{2 \pi}{l_g + l_i}.
\end{equation}
The following sections outline the results of the analyses discussed above.

\section{Dissipation Range Break Frequency and Onset Wavenumber}
\label{sec-kd}

\begin{figure}[]
    \centering
    \includegraphics[width=0.99\textwidth]{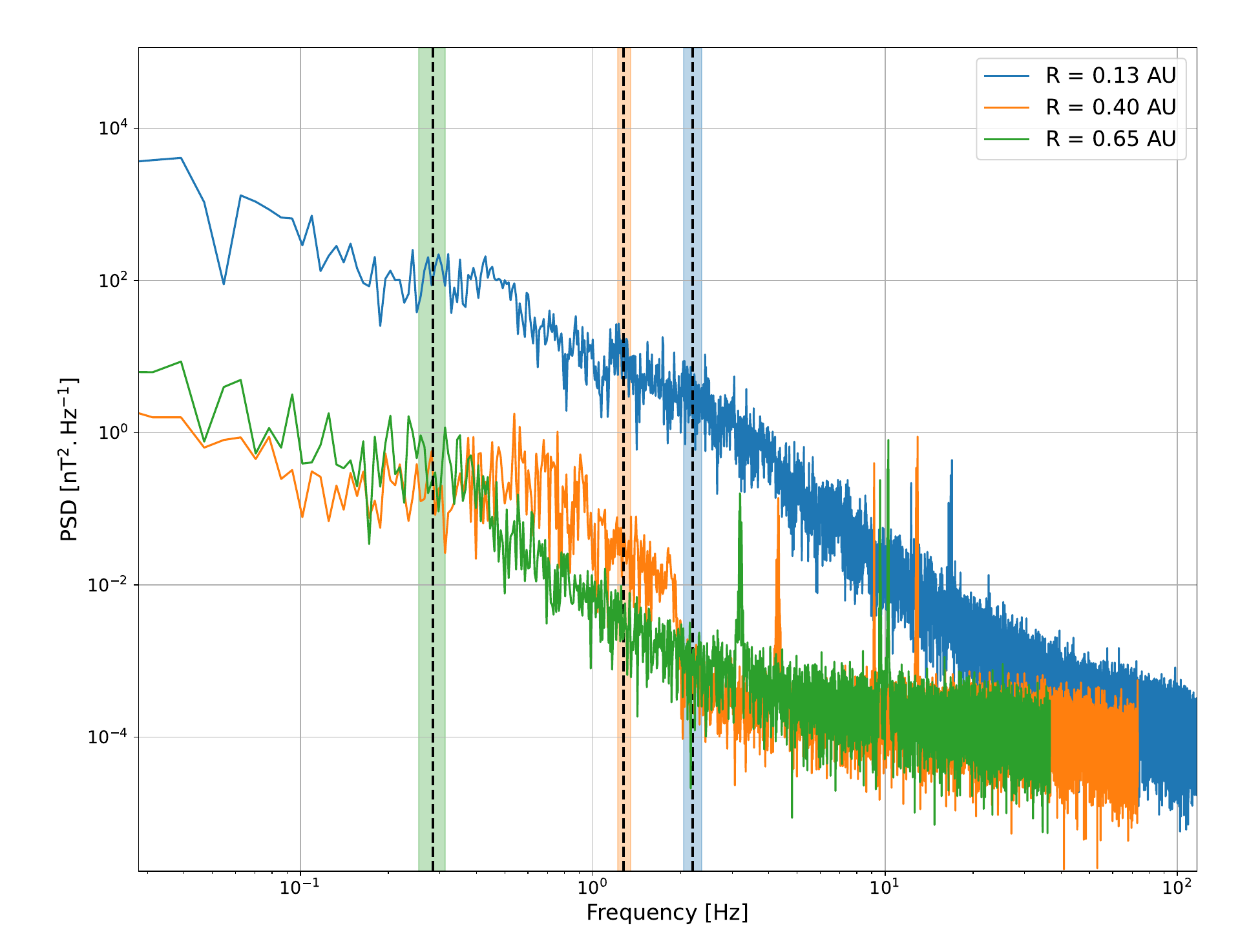}

    \caption{Power spectral distributions at different heliocentric distances. The blue spectrum is from an interval measured at 0.13 au, the orange spectrum from 0.4 au, and the green spectrum as measured at 0.65 au. 
    Vertical dashed lines indicate the estimated break frequency for each spectrum and the shaded regions the error estimate.
    }
    \label{fig:psd}
\end{figure}

\begin{figure}[]
    \centering
    \includegraphics[width=0.99\textwidth]{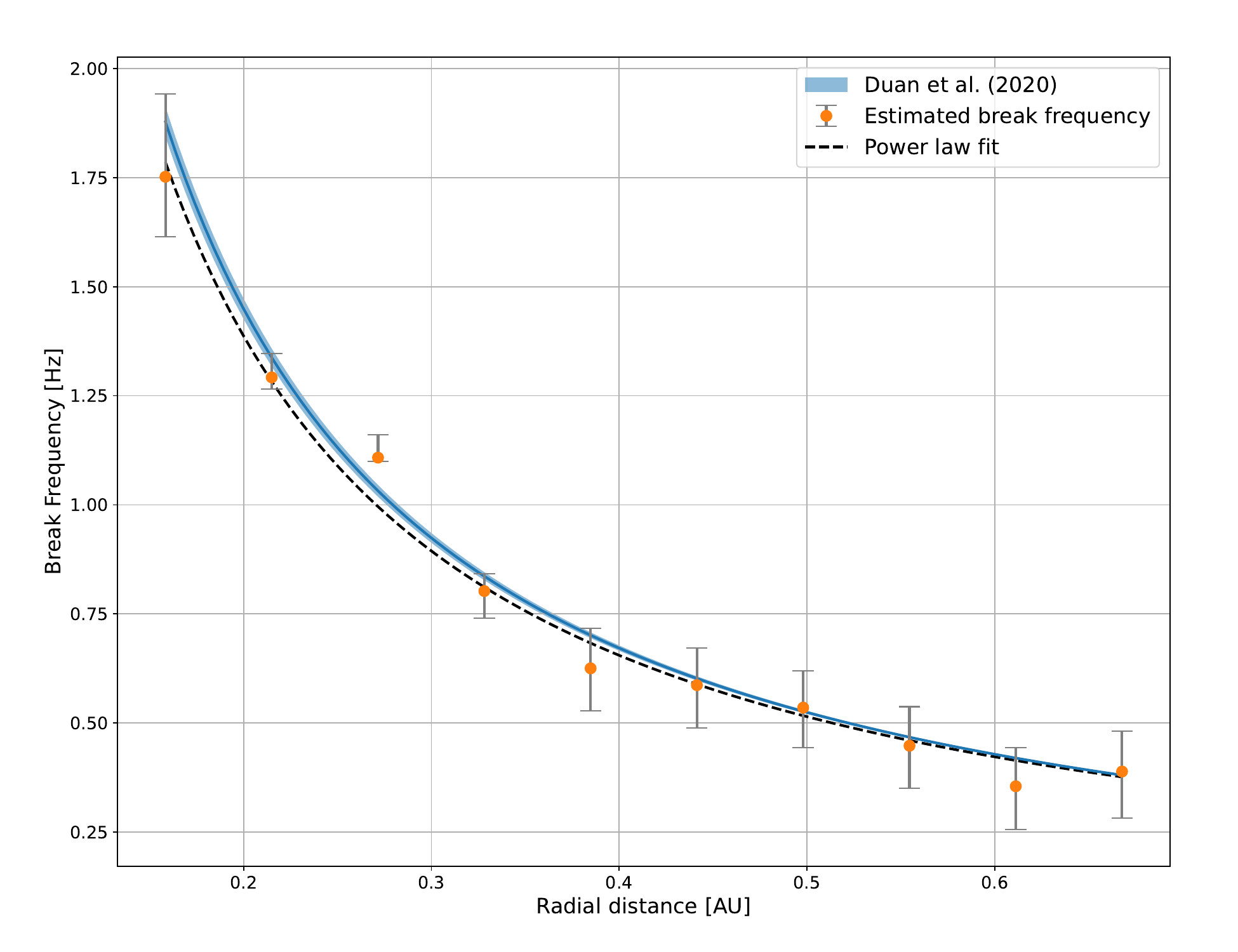}
    \caption{Break frequency, as a function of radial distance, compared with the \citet{Duan2020} power law estimate ($f_b \propto r^{-1.11 \pm 0.01}$) determined from PSP data in the range 0.17 to 0.63 au. The dashed line shows a power law fit to our results ($f_b \propto r^{-1.08 \pm  0.04} $).
    }
    \label{fig:freqrad}
\end{figure}

Figure \ref{fig:psd} shows the magnetic field spectral densities at three different heliocentric distances, 0.13, 0.40 and 0.67 au. 
The dashed vertical lines represent the estimated break frequency and it's estimated error is indicated by the shaded band for each of these examples. 
Sharp peaks in power spectral densities at frequencies above 1 Hz are visible in all of the traces.
These are caused by the spacecraft attitude control system and change in amplitude and frequency slowly over time. 
The increase in total power of the fluctuations and break frequency towards the Sun is expected.
This behaviour can also be seen over wider radial distance in Figure \ref{fig:freqrad}, showing binned estimates of the break frequency as a function of radial distance. 
A power law fit to the data is indicated by a black dashed line ($f_b \propto r^{-1.08 \pm  0.04} $).
This agrees reasonably well with the fit performed by \cite{Duan2020}, shown in blue ($f_b \propto r^{-1.11 \pm 0.01}$).

\begin{figure}
    \centering
    \includegraphics[width=0.75\textwidth]{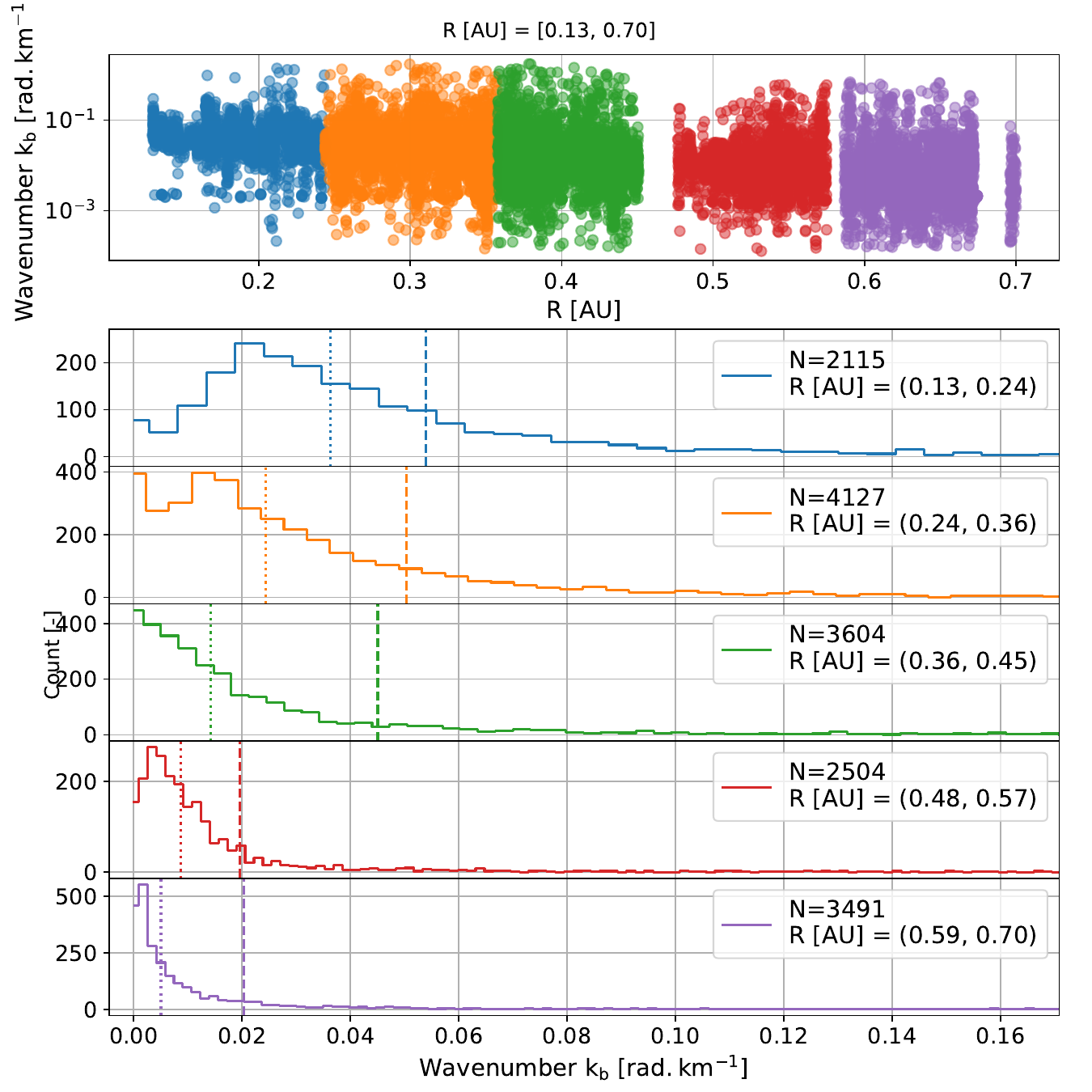}
    \caption{The breakscale wavenumber estimations are binned, according to radial heliocentric distance, into 5 intervals and shown in the top panel, with the wavenumber given in rad/km and radial distance in au. The corresponding distributions obtained for each respective binned dataset are shown in the lower panels as a function of wavenumber. The mean (dashed line) and median (dotted line) of each histogram is indicated by vertical lines.}
    \label{fig:BScale}
\end{figure}

\begin{figure}
    \centering
   \includegraphics[width=0.90\linewidth]{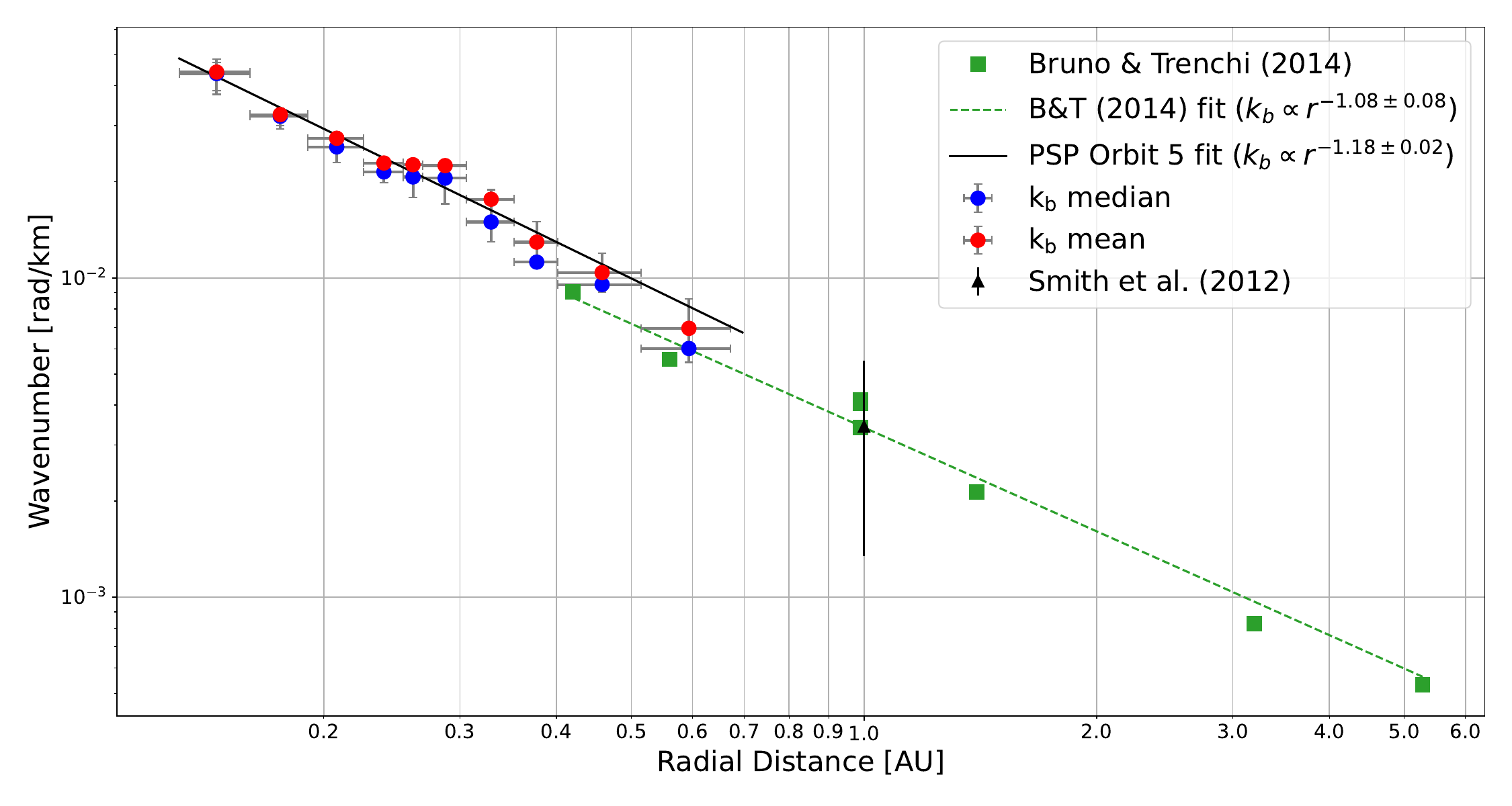}
    \caption{
    The estimated breakscale wavenumber $k_b$, in rad.km$^{-1}$, versus the radial distance for 10 intervals ranging from 0.1 to 0.7 au. The red markers represent the mean $k_b$ values, and the blue markers the median $k_b$ values. The vertical error bars indicate the error derived from the break frequency estimates. Horizontal error bars indicate the radial range covered.
    Comparison is made with $k_b$ reported by \cite{Bruno2014} for radial distances 0.42 to 5.3 au (green squares), as well as the average value for this quantity reported at $1$~au by \cite{SmithEA12} (black triangle), where the error bar indicates the standard deviation of the \cite{SmithEA12} measurements. 
    }
    \label{fig:specBComp}
\end{figure} 

\begin{figure}
    \centering
   \includegraphics[width=0.80\linewidth]{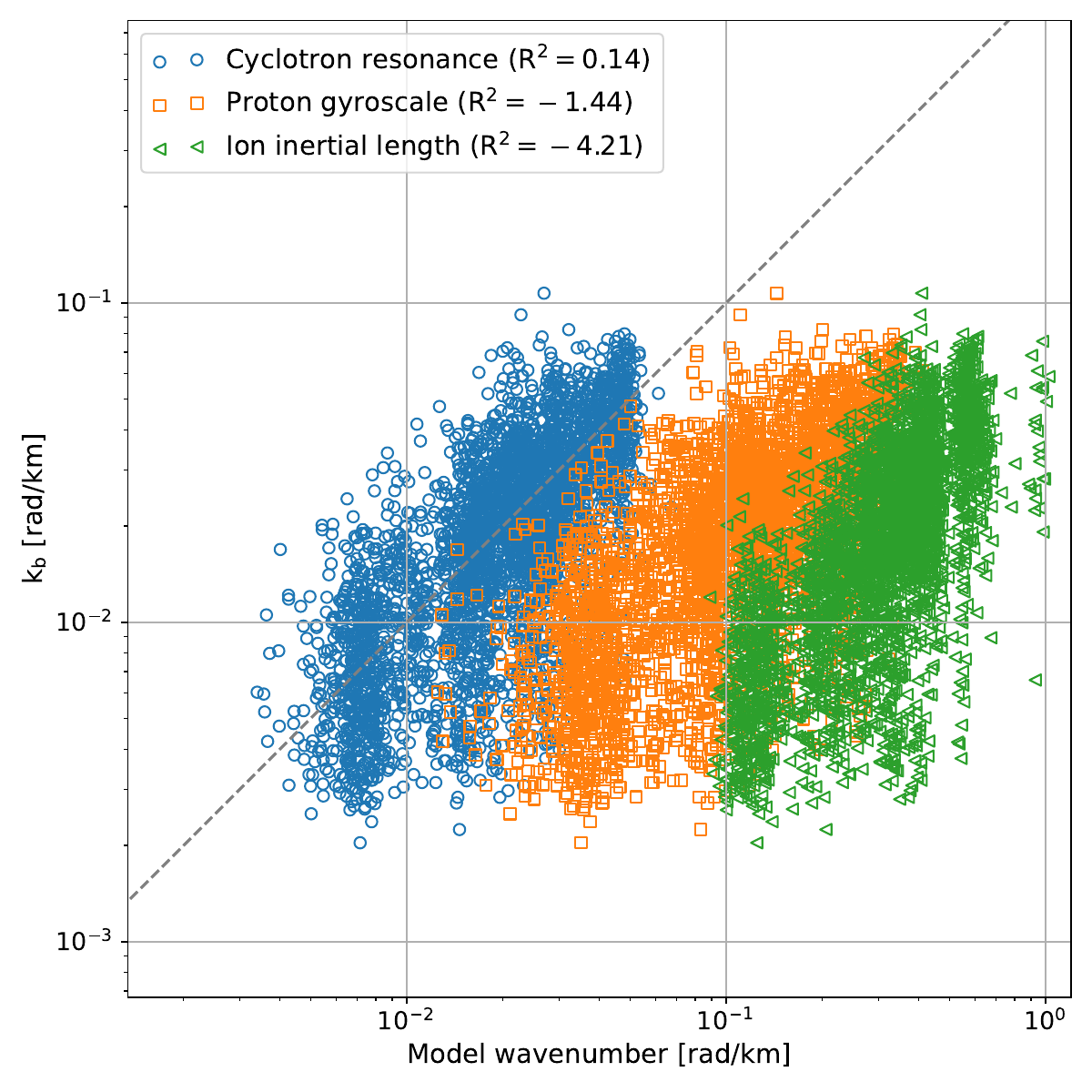}
    \caption{The estimated breakscale wavenumber $k_b$ and a function of the corresponding scale size, calculated from Eqs. \ref{Eq:gyroscale} -- \ref{Eq:cyclotron_resonance}, using {\it in situ} observed plasma data. The dashed line indicates perfect linear correlation with a slope of unity and going through the origin; the coefficient of determination $R^2$ between $k_b$ and the modelled wavenumbers are indicated in the legend.}
    \label{fig:specBComp_insitu}
\end{figure}

The wavenumbers change with radial distance and we show this in Fig.~\ref{fig:BScale}.  
The set of $k_b$ estimates are divided into 5 bins, depicted by the top panel. 
Lower panels show the distribution of $k_b$ for each radial bin. 
As function of radial distance, these distributions become steeper further away from the sun, so that median and mean averages (denoted by short and long dashed lines, respectively) shift further away from distribution peaks at smaller radial distances. 
Overall, distribution peaks shift towards larger values of $k_b$ as radial distances increase, with a corresponding increase in the median and mean average values for this quantity. \\

The PSP results presented in this work enables us to extend earlier estimates of wavenumber to smaller radial distances. 
Figure ~\ref{fig:specBComp} shows $k_b$ corresponding to the median (blue markers) and mean (red markers) values of the distributions illustrated in Fig.~\ref{fig:BScale}.
The green markers correspond to earlier results for this quantity reported by \cite{Bruno2014}, who employ observations from MESSENGER, WIND, and ULYSSES to calculate values of $k_b$ corresponding to an overall radial range spanning 0.42 and 5.3 au. 
The units of these data points are here adjusted for comparison with the values acquired in the present analysis. 
This range overlaps with the range of radial distances considered here and, as can be seen from Fig.~\ref{fig:specBComp}, the break wavenumbers calculated by \cite{Bruno2014} tie in reasonably well with estimates from the present study. 
An average of the values for $k_b$ reported by \citet{SmithEA12} from an analysis of spacecraft observations at $1$~au is also shown, and estimates the study of \citet{Bruno2014} fall well within the uncertainty of that value. 
Not unexpectedly, the radial behaviour of the observations shown in Fig.~\ref{fig:specBComp} suggests a power law radial dependence for $k_b$, similar to that seen for the break frequency. 
Accordingly, the figure also shows a power law fit to the observations, with exponent $-1.18\pm 0.02$, a value steeper than that reported by \citet{Bruno2014}.

In order to directly compare the dissipation range onset wavenumbers calculated here with those wavenumbers corresponding to the various lengthscales that have been proposed previously, Fig.~\ref{fig:specBComp_insitu} shows $k_b$ as function of wavenumbers corresponding to the cyclotron resonance scale, the proton gyroscale, and the ion inertial length, as calculated from the corresponding {\it in situ} measurements, with corresponding correlation coefficients indicated in the legend. 
Of all three lengthscales, the cyclotron resonance wavenumber best fits the dissipation range onset wavenumber calculated here, although the considerable scatter in the data points leads to a relatively coefficient of determination ($R^2$). 

The dashed line in the figure indicates where $k_b$ is equal to the model wavenumber. When using the cyclotron wavelength model wavenumber for comparison (blue circles), the results follow this trend extremely well.\\

\section{Spectral indices}
\label{sec-si}

\begin{figure}
    \centering
    \includegraphics[width=0.49\textwidth]{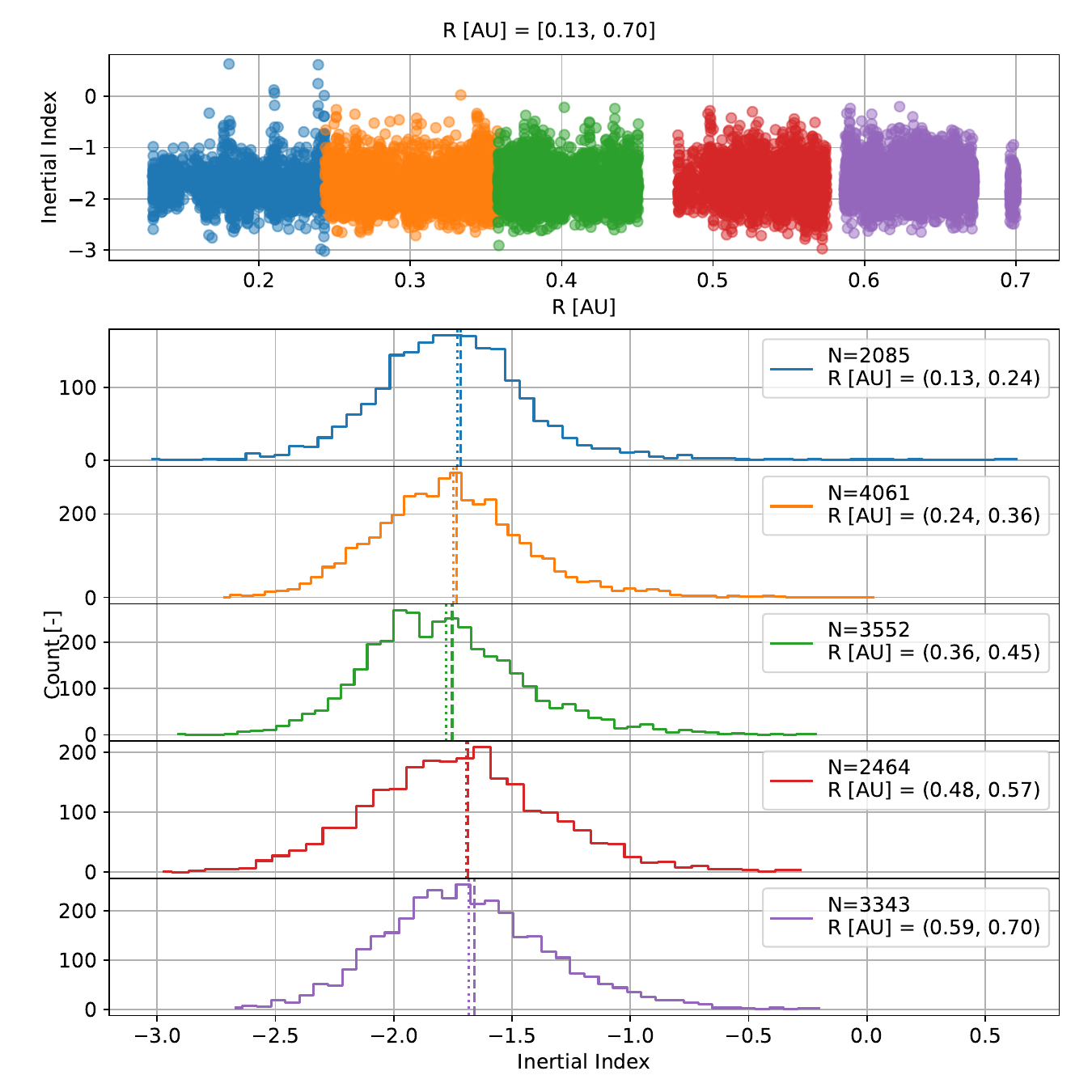}
    \includegraphics[width=0.49\textwidth]{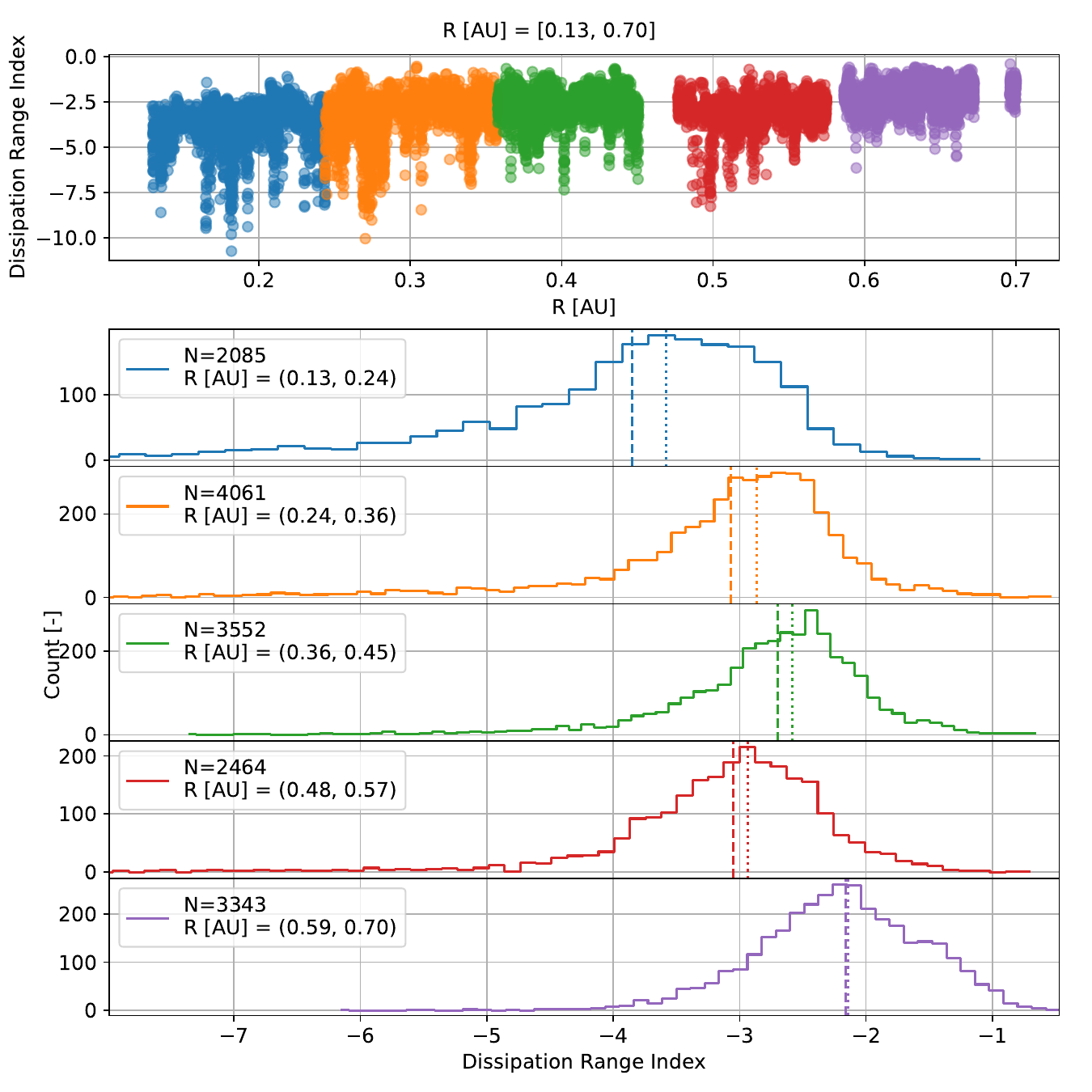}
    \caption{Similar to Fig. \ref{fig:BScale}, but now the value of the inertial range power law index (left column) and dissipation range power law index (right column) are binned into several radial intervals.}
    \label{fig:inertI}
\end{figure}


\begin{figure}
    \centering
    \includegraphics[width=0.90\textwidth]{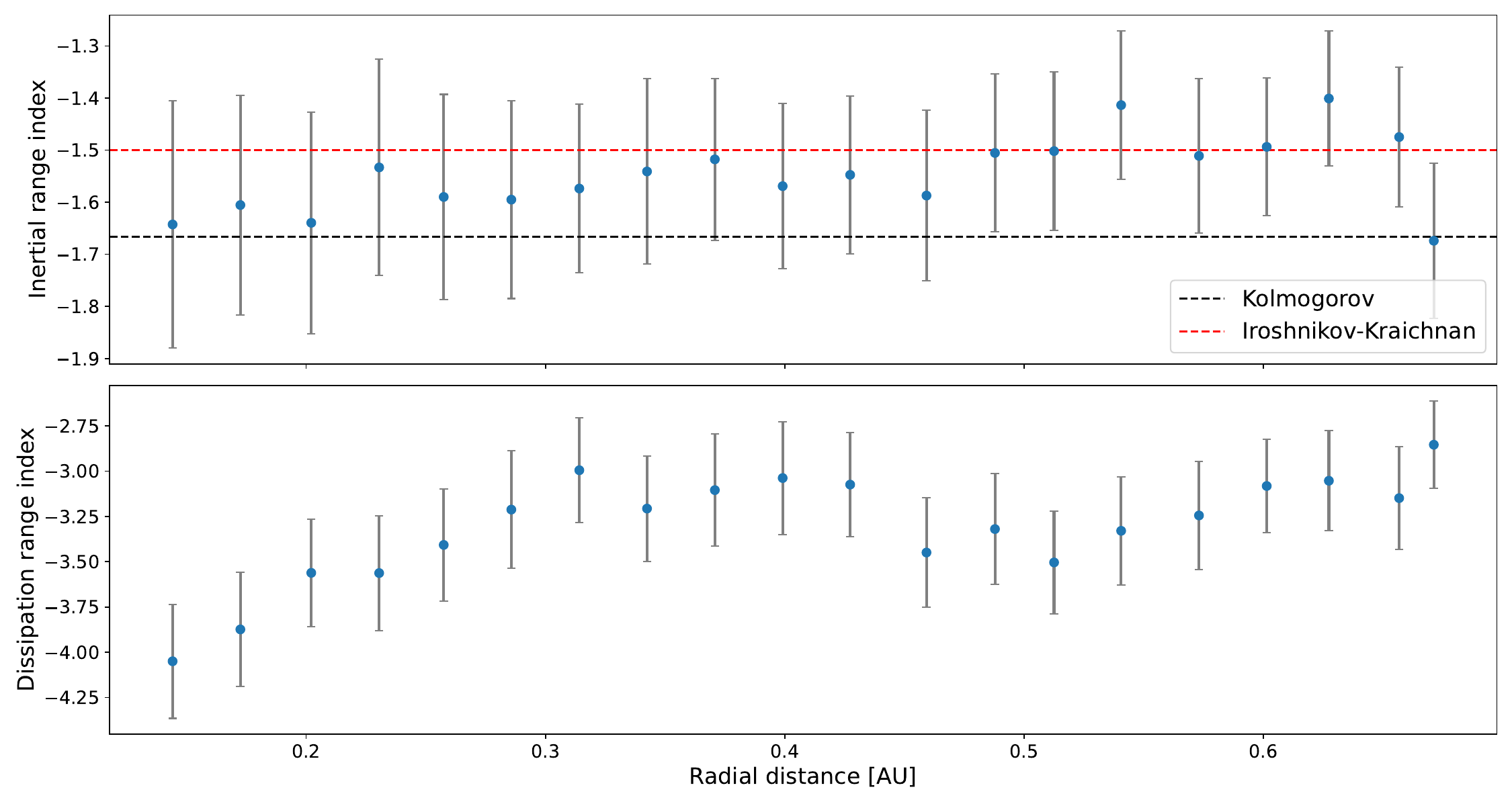}
    \caption{Computed inertial (top) and dissipation range (bottom) power law indices as a function of radial distance.}
    \label{fig:indices}
\end{figure}

\begin{figure}
    \centering
    \includegraphics[width=0.90\textwidth]{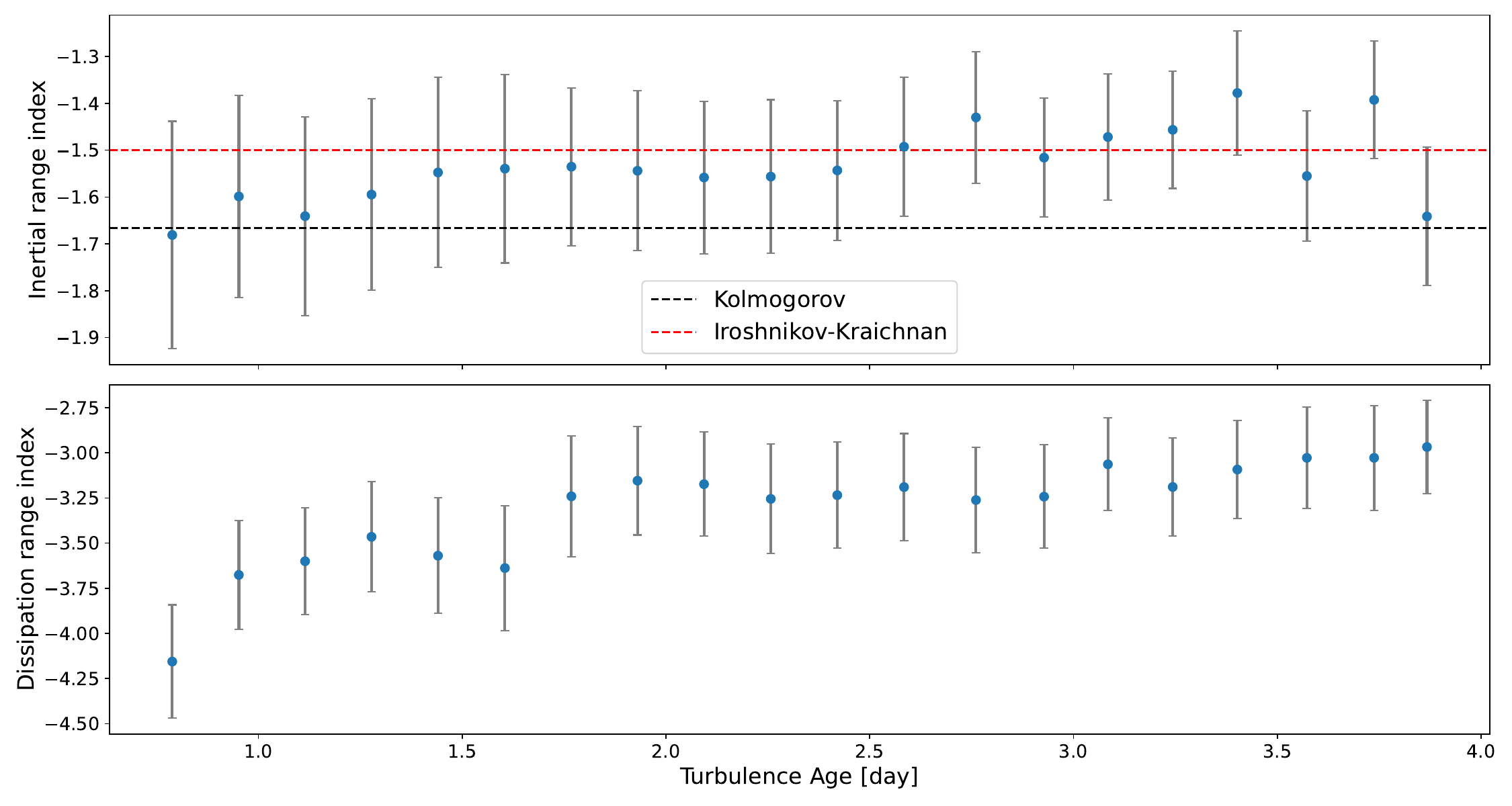}
    
    \caption{The inertial (top) and dissipation range (bottom) power law indices as a function of solar wind age.}
    \label{fig:indices2}
\end{figure} 

In a similar method to that use in the previous section, Fig. \ref{fig:inertI} shows the estimated power law indices of the inertial and dissipation range turbulence, along with their distributions, binned into different radial intervals. The median and standard deviation of each binned histogram is calculated and shown in Fig. \ref{fig:indices} as a function of radial distance. The inertial range index distribution does not show a change in width, and remains relatively constant with an increase in radial distance. The dissipation range index shows a constant distribution width, and generally decreases as radial distances get larger. This is shown more explicitly in Figure \ref{fig:indices}, where both spectral indices, with their accompanying uncertainties, are shown as function of the average radial distances corresponding to the bins for which they were calculated. The inertial range spectral indices appear to remain relatively constant as function of radial distance (within uncertainty). Furthermore, it is not clear whether either the Kolmogorov or Iroshnikov-Kraichnan values for this quantity, indicated respectively by the black and red lines on the figure, is favoured, once more due to the range of uncertainty. Although this result is similar to the inertial range spectral indices calculated for the 5th and 7th PSP orbits calculated by \citet{ZhaoEA22}, there are some differences, as the averaged indices reported by those authors steepen from a value roughly between the Iroshnikov-Kraichnan and Kolmogorov values, to a value approximately equal to the Kolmogorov index beyond $\sim 0.35$~au for the 5th orbit data \cite[see also][]{SioulasEA22}.

Dissipation range spectral indices, however, display a clear radial dependence, increasing in absolute value as radial distance increases, similar to what is reported by \citet{franci20}. {It is interesting to note that, were the violation of the Taylor hypothesis to play a significant role, a flatter dissipation range spectrum would be expected \citep{KleinEA14}. The radial behaviour of this quantity} may be related to the increase in power of the turbulence with decreasing radial distance (see Figure~\ref{fig:psd}), in qualitative agreement with what was reported by \cite{smith06} in their analysis of dissipation range spectral indices at $1$~au, by \citet{BrunoEA14}, who found a correlation between steeper dissipation ranges, and enhanced turbulence levels, and by \citet{HuangEA21} in their analysis of first orbit PSP data. The radial decrease in the dissipation range spectra index is, however, not uniform, as indicated by the slight drop in this quantity between $\sim 0.45$~au and $\sim 0.6$~au in Fig. \ref{fig:indices}. This behaviour corresponds to a marked increase in the solar wind speed shown in Fig.~\ref{fig:PSP_plasma_overview}, with the implication that the behaviour of the dissipation range spectral index at these radial distances may be a reflection of the behaviour of this quantity at smaller radial distances due to the fact that the solar wind is `younger' here. To investigate this, the age of the solar wind $\tau = r/V_{sw}$ corresponding to each point in Fig. \ref{fig:indices} was calculated, and the spectral indices were plotted as function of this age in Fig.~\ref{fig:indices2}. The inertial range spectral indices behave in a relatively uniform manner as function of solar wind age, while the dissipation range spectral indices clearly become steeper as solar wind age decreases.\\

\section{Discussion and conclusions}
\label{sec-discussion}

This work extends the study of \citet{Duan2020}, by considering the radial evolution of the spectral break between the inertial and dissipation ranges of turbulence power spectra calculated for PSP data taken during the 5th perihelion of that spacecraft, as well as the spectral indices associated with these ranges. The radial dependence of the spectral break reported on here closely resembles that found by \citet{Duan2020} for the the cruise phase of the 2nd PSP orbit, as well as tying up with that reported at larger radial distances by \citet{Bruno2014}, as well as an average value for this quantity calculated from the results of \citet{SmithEA12}. We find a radial dependence of $k_b \propto r^{-1.18 \pm 0.02}$. Furthermore, a comparison of the break frequencies calculated here with frequencies corresponding to the proton gyroradius, ion inertial length, and the cyclotron resonance scale, all computed using {in situ} PSP observations of the various plasma quantities these scales depend on, found that these break frequencies correspond most closely with those corresponding to the cyclotron resonance scale. This finding is in agreement with that of \citet{WoodhamEA18} in their analysis of Wind data. These results provide a valuable benchmark against which the results of various turbulence transport models \citep[e.g.][]{Eng2018,AdhikariEA21} can be tested, as well as a valuable input for solar energetic particle and cosmic ray transport models. 

The present study finds, in contrast to previous studies \cite[e.g.][]{chenetal2020,ShiEA21,ZhaoEA22,SioulasEA22} that the inertial range spectral index, within uncertainties, remains relatively constant as function of heliocentric radial distance. However, this index does steepen for intervals of greater age, corresponding to relatively slower solar wind speeds. This discrepancy with the results of prior studies may be due to the fact that the present study does not distinguish between intervals of greater or lesser Alfv\'enicity. The ambiguity in the results presented here may then be due to the fact that that inertial range spectral indices have been observed to be steeper for intervals of low Alfv\'enic content (characterized by \citet{SioulasEA22} as intervals of low normalized cross-helicity), and vice versa \cite[see also][]{ShiEA21}. However, the steepening of inertial range spectral indices reported here for older turbulence intervals does agree with the findings of \citet{SioulasEA22}, who report such a steepening for slow solar wind intervals.

The dissipation range spectral indices reported on here show a clear radial dependence, becoming less steep at larger radial distances, in agreement with previous studies \cite[e.g.][]{franci20}. This could be related to the increase in turbulence levels closer to the Sun \cite[e.g.][and references therein]{AdhikariEA21b,ZankEA21}, and would be in accordance with a correlation between enhanced turbulence levels at $1$~au and steeper dissipation range spectral indices reported by \citet{smith06}. It is interesting to note that, when intervals are binned according to the age of the turbulence, dissipation range spectral indices almost uniformly decrease as radial distance increases. 

{Another possibility for the mechanism that steepens the dissipation range spectral index is the presence of Ion Cyclotron Waves (ICWs). \cite{BowenEA20} showed that the number, amplitude and duration of ICW packets increases closer to the Sun. The presence of ICWs causes a bump in the spectrum at the proton cyclotron resonance scale \citep{Lionetal2016, WicksEA16, WoodhamEA18, TelloniEA19} and so can appear as a steepening of the dissipation range spectrum. Our method rejects spectra with large standard deviation of the fitted lines and so spectra with large peaks due to ICWs will be rejected, but we cannot rule out that low-amplitude ICW effects on the spectra have not been measured as steepening of the dissipation range.}

Future work aims to extend the current analysis in two ways. Firstly, by considering longer data intervals, so as to include a portion of the energy-containing range of the turbulence power spectrum, thereby allowing for the calculation of the inertial range outerscale. This, in turn will enable us to include the disruption scale \citep[e.g.][and references therein]{TerresLi22} in our comparative analyses. Secondly, the analysis will be extended to other PSP perihelia, taking into account various additional factors, such as solar wind speed and plasma-$\beta$, that are known from previous studies \citep[e.g.][]{Chenetal2014,WangEA18,SioulasEA22} to influence the dissipation range spectral break frequency. {It should also be noted that the influence of Taylor's hypothesis on results calculated for the dispersive regime should be investigated, in the manner proposed by e.g. \citet{BP19}, in more detail.} Future measurements with the MeerKAT radio telescope and the Square Kilometer Array (SKA) are also planned to get information about solar wind density fluctuations at very small scales inside the Alfv\'en radius, and thus close the gap between the Sun and $\sim$10 $R_{\odot}$.

Software repository at \url{https://bitbucket.org/stefansansa/pspanalysis/}.

\acknowledgments

This work is based on the research supported in part by the National Research Foundation of South Africa (NRF grant numbers 119424, 120345, 120847, and 137793). Opinions expressed and conclusions arrived at are those of the authors and are not necessarily to be attributed to the NRF. The responsibility of the contents of this work is with the authors. Figures prepared with Matplotlib \citep{hunter} and certain calculations done with NumPy \citep{harrisetal2020}. RTW is funded by STFC Grant ST/V$006320$/$1$.
EPK was supported by STFC grant ST/T000422/1. Parker Solar Probe was designed, built, and is now operated by the Johns Hopkins Applied Physics Laboratory as part of NASA’s Living with a Star (LWS) program (contract NNN$06$AA$01$C). Support from the LWS management and technical team has played a critical role in the success of the Parker Solar Probe mission. Thanks to the FIELDS team for providing data (PI: Stuart D. Bale, UC Berkeley). Thanks to the Solar Wind Electrons, Alphas, and Protons (SWEAP) team for providing data (PI: Justin Kasper, BWX Technologies).



\bibliography{agusample.bib}

\end{document}